\documentclass[12pt]{iopart}
\usepackage{graphicx}
\usepackage{harvard}
\usepackage{url}
\usepackage{iopams}  
\renewcommand{\harvardurl}{URL: \url}
\begin{document}

\title[Empirical potential structure refinement on PTFE and PCTFE]{Empirical Potential Structure Refinement of semi-crystalline polymer systems: polytetrafluoroethylene and polychlorotrifluoroethylene}

\author{A K Soper$^1$, K Page$^2$ and A Llobet$^2$}

\address{$^1$ STFC Rutherford Appleton Laboratory, Harwell Oxford, Didcot, Oxon, OX11 0QX, UK}
\address{$^2$ Lujan Neutron Scattering Center, Los Alamos National Laboratory, MSH805, Los Alamos, NM  87545, USA}

\ead{alan.soper@stfc.ac.uk}
\begin{abstract}
Empirical potential structure refinement (EPSR) simulations are performed on total neutron scattering data from powder samples of polytetrafluoroethylene (PTFE) and polychlorotrifluoroethylene (PCTFE), both at 300K. Starting from single strands of polymer consisting of between 30 and 60 monomers of tetrafluoroethylene and chlorotrifluoroethylene in each case, hexagonal simulation cells are constructed consisting of an array 25 (5 $\times$ 5) such strands placed on a hexagonal lattice. Allowed simulation moves are polymer translation moves along all three Cartesian axes, whole polymer rotations about the polymer axis, and individual atom moves within each polymer. For PTFE a number of Bragg peaks are visible in the scattering data and these are found to be consistent with a lattice spacing $a(=b)=5.69(1)$\AA\ with a dihedral angle along the (helical) chain of 166$^\circ$ which gives a repeat distance along the chain ($c$-axis) of $\sim$19.6\AA. The positions of the Bragg peaks are well reproduced by this model, although there is a mismatch in the amplitudes of some of the higher order reflections between simulation and data. For PCTFE there is only one visible Bragg peak (100) which is well reproduced by a hexagonal lattice of atactic parallel polymers with a spacing of $a(=b)=6.37(1)$\AA. In this case the absence of distinct reflections along the polymer $c$-axis makes characterisation of the internal dihedral angle difficult, but a model with a dihedral angle of 166$^\circ$ was less successful at fitting the diffuse scattering than a model where this angle was set to 180$^\circ$, giving a nearly straight \textit{trans} (zig-zag) structure. For PCTFE little change in structure could be discerned when the material was heated to 550K, apart from a slight increase in lattice spacing. In both cases there is substantial diffuse scattering between the Bragg peaks, and this is correctly replicated by the EPSR simulations.
\end{abstract}

\submitto{\JPCM}
\maketitle

\section{Introduction}
In a landmark series of papers spanning nearly two decades, G. R. Mitchell and coworkers have developed a reverse Monte Carlo method for modelling the structure of polymers based on neutron total scattering data \cite{mitchell1992:1,mitchell1994:1,mitchell1995:3,mitchell1995:2,mitchell1995:1,mitchell2011:1}. Reverse Monte Carlo (RMC) was originally developed as a method for modelling atomistic structures using the scattering data from liquids and glasses as a constraint instead of the more usual force fields of conventional computer simulations \cite{mcgreevypusztai}. The advance made by Mitchell was to introduce the necessarily restrictive atomic positions which arise in a polymer from intra-chain bonding, so that bond lengths, bond angles, and dihedral angles all had to be defined in order for the model to be built. Initial determination of these parameters was sometimes achieved using quantum chemistry calculations \cite{mitchell1994:1}, with final refinement made against the scattering data. The method consisted of building multiple chains using a range of bonding parameters, then gradually refining those parameters to give a model that most closely fit the scattering data. This process was therefore somewhat distinct from the original RMC method which consisted of individual atom moves and constraining these moves to give the best fit. Here the bonding parameters were adjusted to give the best fit. As a result a plethora of information about the likely local and longer range structure of polymer was obtained from the scattering data, on the basis of relatively few initial assumptions about bond lenths, bond angles and dihedral angles \cite{mitchell2011:1}. The Mitchell method was subsequently complemented by more traditional RMC work \cite{rmc1998:1,rmc2004:1}, in which the polymer chain was built by means of a self-avoiding random walk. Individual atomic moves were allowed, but intra-chain bonds and bond angles were controlled by a series of bonding constraints which prevented neighbouring atom pairs moving outside a specified range of intra-chain distances and angles. A useful view of local structure in polymers in this period is given by \cite{winokur1998local}.

These studies were applied to cases of liquid or amorphous polymers where the scattering is diffuse at all $Q$ values. In addition, the different methods build a single polymer chain within the simulation box and the inter-chain correlations of the real material are modelled by the intra-chain correlations over the medium or longer distance range. This is reasonable in a liquid or amorphous structure, since the scattering pattern in such cases is dominated by intra-chain correlations at short distances. However when the polymer crystallises, even if the crystallisation is only partial, a new set of constraints arise, based precisely on the nature of the inter-chain packing, which in turn arises from the crystalline long range order that is now imposed on the material. Yet even when crystalline packing of the chains becomes apparent, there is likely to remain a significant degree of disorder along the chains arising from the distribution of bond lengths, bond angles and dihedral angles, and the longer range fluctuations in polymer geometry that these distributions induce. There may also be some positional disorder between chains to a greater or lesser extent. 

To model such structures requires a different approach compared to simply selecting bonding criteria since if the endeavour is to have any chance of success on a realistic timescale it is essential to build in as much knowledge about the crystal symmetry and packing that is known either beforehand or by inspection of the data themselves. The structure refinement problem is one of dealing with a marked degree of crystallinity combined with significant diffuse scattering which can occur over many length scales: we have to generate a mixture of disorder over a range of length scales and combine this with the long range order of the crystal lattice. As a result the way the polymer molecules are packed certainly cannot be ignored in this case.

For this first study of crystalline polymers using total scattering techniques combined with computer simulation structure refinement we have chosen the relatively stiff polymer, polytetrafluoroethylene, PTFE, and its chlorinated variant, polychlorotrifluoroethylene, PCTFE. PTFE has a rather complex phase diagram, \cite{crystalstruct2006:1}, existing in at least three phases in the temperature range 273-313K at ambient pressure. Less appears to be known about the phase behaviour of PCTFE, but in both cases crystallographic studies suggest a hexagonal packing of long linear filaments of the polymer, with a slight twist along each of the C-C bonds which gives rise to a helical structure along the chain instead of a simple linear zig-zag chain \cite{crystalstruct1954:1,crystalstruct1973:1}. This system is therefore readily amenable to setting up a crystalline simulation model. 

For the work described here we have adapted the empirical potential structure refinement (EPSR) method, \cite{epsrrecs2001:1,epsrrecs2005:1,epsrrecs2012:1} to both generate the polymer structure initially, assemble the polymer strands into a crystalline array, and then structure refine both the internal structure of the individual polymer strands as well as the inter-chain correlations. Although originally derived from RMC, EPSR is distinct in that the atomic and molecular moves are driven entirely by forces. These forces take several forms: harmonic forces to control short range intra-molecular distances, repulsive and dispersion forces to control packing between different molecules and to prevent unphysical atomic overlap, and an empirical force field derived from a comparison of the simulated structure factor with the experimental structure factor. It is the latter force that introduces the scattering data into the structure refinement process. In the following sections we illustrate the application of this enhanced EPSR method to the structure refinement of the crystalline or semi-crystalline polymers, PTFE and PCTFE, at 300K, and for the latter polymer examine whether any structural transition is visible at 550K

\section{Methods}
Total neutron scattering data on the two polymers were collected at the Lujan Center at Los Alamos Neutron Science Center (LANSCE) on the High Intensity Powder Diffractometer (HIPD) and the Neutron Powder Diffractometer (NPDF) \cite{savedrecs2002:1}. Experiments were completed utilizing vanadium canister sample holders with approximately 1 cc of material. The materials were prepared in the form of a powder of small flakes to produce a sample appropriate for total neutron scattering measurements. Teflon data were collected on the high flux instrument HIPD at 300 K for approximately 5 hours, while Kel-F data were measured at 300 K, 350 K, and 550 K for approximately 5 hours each.  A high resolution data set was collected for Kel-F on the NPDF instrument at 300 K for approximately 9 hours.  

In all cases data reduction was completed with the software package PDFgetN \cite{savedrecs2000:1}.  The measured data were corrected to account for background scattering, container scattering, as well as the incident spectrum.  Sample data were corrected for absorption and multiple scattering, and normalized to produce the total scattering factor, $S(Q)$, corresponding to the PDFFIT formalism \cite{savedrecs2001:1}. Here $S(Q)$ is defined as
\begin{equation}
S(Q)-1={\sum_{\alpha\beta\geq\alpha}c_{\alpha}c_{\beta}\langle b_{\alpha}\rangle\langle b_{\beta}\rangle(2-\delta_{\alpha\beta})\left[A_{\alpha\beta}(Q)-1\right]}/{\left(\sum_{\alpha}c_{\alpha}\langle b_{\alpha}\rangle\right)^2}
\label{sofq}
\end{equation}
where $c_{\alpha}$ is the atomic fraction and $\langle b_{\alpha}\rangle$ is the spin and isotope averaged neutron scattering length for atomic species $\alpha$, and the partial structure factors are defined according to
\begin{equation}
A_{\alpha\beta}(Q)-1=4\pi\rho_0\int_{0}^{\infty}r^{2}\left(g_{\alpha\beta}(r)-1\right)\frac{\sin Qr}{Qr}dr
\label{aofq}
\end{equation}
with $\rho_0$ the atomic number density of the material, $g_{\alpha\beta}(r)$ the site-site radial distribution function for the atom pair ($\alpha,\beta$), and $Q=4\pi\sin \theta/\lambda$ is the wave vector change in the scattering experiment, with $2\theta$ the detector scattering angle and $\lambda$ the neutron wavelength. In a time-of-flight experiment such as here the detector scattering angle is held fixed and the incident neutron wavelength varied, although for both HIPD and NPDF there are large arrays of detectors at different scattering angles. Final data normalisation was achieved by adjusting the effective sample density to obtain the correct limiting behaviour $S(Q) = 1$ at high $Q$. 

Also of interest is the Fourier transform of the structure factor:
\begin{eqnarray}
G(r)-1&=&\frac{1}{2\pi^2\rho_0}\int_{0}^{\infty}Q^{2}\left(S(Q)-1\right)\frac{\sin Qr}{Qr}dr \nonumber \\
&=&{\sum_{\alpha\beta\geq\alpha}c_{\alpha}c_{\beta}\langle b_{\alpha}\rangle\langle b_{\beta}\rangle(2-\delta_{\alpha\beta})\left[g_{\alpha\beta}(r)-1\right]}/{\left(\sum_{\alpha}c_{\alpha}\langle b_{\alpha}\rangle\right)^2}
\label{biggofr}
\end{eqnarray}
This total pair distribution function (PDF) is helpful for assessing the quality of fit in real ($r$) space.

\section{Empirical potential structure refinement simulations (EPSR)}

The basic EPSR method has been described in several publications \cite{epsrrecs2001:1,epsrrecs2005:1,epsrrecs2012:1} so that information will not be repeated here. In order to model these polymeric structures some new commands were introduced into the EPSR package. In particular the \textbf{dockato} command allows fragments of molecules to be built into larger fragments or monomers, and then these monomers can be combined into longer polymers, with the dihedral angle at the point of joining specified if required. The \textbf{dockato} program can be used in the same manner as described in \cite{rmc1998:1} to build a self-avoiding chain if required. Having built the polymer, the \textbf{alignmole} command allows the polymer to be aligned along one of the Cartesian axes, in this case the $z$-axis which is coincident with the crystallographic $\textbf{c}$-axis. The single strand of polymer is placed in a hexagonal unit cell, with the angle between the $\textbf{a}$ and $\textbf{b}$ axes set to 120$^\circ$ and with equal magnitudes for these two axes. From the single unit cell a 5$\times$5 array of unit cells is constructed, using the command \textbf{makelatticeato}. Finally the polymers are given a realistic degree of atom-atom disorder using the standard command \textbf{fmole} - this is to emulate the zero point disorder that would occur in any atomistic representation of a molecule. Full details of these new commands will appear in the EPSR User Manual, \cite{epsrrecs2013:1}.

\subsection{Calculating Bragg peak intensities}

Another new feature of the present method is that because it is assumed the underlying structure is crystalline, the Bragg peak intensities need to be calculated separately from the local, diffuse scattering: this is the only practical way to capture these sharp peaks without resorting to a model with a very large number of unit cells, which would be cumbersome and slow to simulate. Hence in the present instance the Bragg scattering amplitude is calculated using the formula \cite{kittel1971}
\begin{equation}
B(hkl)=\sum_{j} b_{j} \exp\left[ -2\pi i\left(hx_{j} +ky_{j}+lz_{j} \right)\right] \label{bhkl}
\end{equation}
where the sum is over all the atoms in the simulation box, $b_{j}$ is the scattering length of atom $j$, $(hkl)$ are the standard crystallographic indices and $(x_{j},y_{j},z_{j})$ are the fractional coordinates of the $j$th atom along the $\textbf{a,b,c}$ lattice vectors respectively. No restriction is placed on the choice of $(hkl)$ values except that in the present instance where the simulation box is a supercell made up of 5 unit cells along each of the $\textbf{a}$ and $\textbf{b}$ axes, only $(hk)$ values which are multiples of 5 are used. For example a Bragg peak listed as (110) would actually be calculated as (550) within the EPSR simulation. Also in the present case, no equivalent multiplicity is assumed along the $\textbf{c}$-axis (polymer axis) at the outset, due to the indefinite length of the real polymer strands, although, as will be seen below, for PTFE, a multiplicity does emerge naturally from the choice of dihedral angle along the polymer chain when attempting to create the Bragg reflections at the observed positions.

Equation (\ref{bhkl}) is sometimes referred to as the crystallographic structure factor but of course has to be carefully distinguished from the total scattering structure factor (\ref{sofq}). The Bragg scattering intensities are given by
\begin{equation}
I_{tot}(hkl)=B(hkl)B^{*}(hkl) \label{ihkl}
\end{equation}
which can be separated into individual site-site terms as in (\ref{sofq}) if required. However this definition will include the single atom scattering, and will represent the total scattered intensity from the entire simulation box, whereas in (\ref{sofq}) scattering intensities are usually normalised per scattering atom, and the calculation is for a limited range of $r$, normally set by the half the simulation box dimension \cite{allentildesley}.

Bragg reflections occur at discrete Q values:
\begin{equation}
\textbf{Q}_{hkl}=h\widehat{\textbf{a}}+k\widehat{\textbf{b}}+l\widehat{\textbf{c}} \label{qhkl}
\end{equation}
where $\widehat{\textbf{a}},\widehat{\textbf{b}},\widehat{\textbf{c}}$ are the reciprocal lattice vectors. In order to include the Bragg peaks in the calculation of the total scattering structure factor it is necessary to broaden them with a function which represents the resolution of the measurements as well as sample effects such as particle size broadening. In the present instance no attempt has been made to generate a precise description of the Bragg peak shape function. Instead the Bragg spots in reciprocal space are broadened by a 3-dimensional Lorentzian function which has the form
\begin{equation}
P(Q,Q_{hkl})=\frac{\widehat{V_B}}{\pi^{2}\gamma}\left\lbrace\frac{1}{[\gamma^2+(Q_{hkl}-Q)^2]}-\frac{1}{[\gamma^2+(Q_{hkl}+Q)^2]} \right\rbrace \label{pofqhkl}
\end{equation}
where $\gamma$ is a width in reciprocal space to be specified by the user (it can be determined approximately from the shape of the main Bragg peaks for example) and $\widehat{V_B}=\widehat{\textbf{a}}\centerdot\widehat{\textbf{b}}\wedge\widehat{\textbf{c}}$ is the volume of the reciprocal lattice. 

This function is used to give a simulated scattering intensity continuous in $Q$. In order to put the calculated Bragg intensities onto the same scale as the total scattering structure factor, it is necessary to smear the Bragg peaks with this function, subtract the single atom scattering, divide by the total number atoms in the simulation box, $N_{S}$, and divide by the mean scattering length squared as in (\ref{sofq}) to give a total Bragg interference function or structure factor:
\begin{equation}
S_{\mathrm{Bragg,int}}(Q)=\frac{\sum_{hkl}I_{tot}(hkl)P(Q,Q_{hkl})-\sum_{j} b^{2}_{j}}{N_{S}\left(\sum_{\alpha}c_{\alpha}\langle b_{\alpha}\rangle\right)^2} \label{sbraggint}
\end{equation}

In practice the calculation of Bragg peak intensities is efficient at low $Q$ where the number of Bragg peaks is small, but rapidly becomes inefficient at high $Q$ if one or more of the lattice vectors is large. Therefore a maximum $Q$ is imposed for calculating the Bragg structure factor (\ref{sbraggint}) where significant Bragg intensities are no longer visible in the data. In the present work a total of $\sim 11,200$ Bragg peaks were calculated, giving a maximum $Q$ for the Bragg peak calculation of $\sim 6.5$\AA$^{-1}$. This number or peaks was required by the large $\textbf{c}$ axis in these simulations which extended up to 156.8\AA\ for the longest polymer chains simulated. 

\subsection{Calculating the structure factor from the simulation}

There are therefore two alternative routes to calculating the structure factor from the computer simulation. The first method is to calculate the site-site distributions, $g_{\alpha\beta}(r)$,  directly from the simulation box, then Fourier transform this to $Q$ space using equations (\ref{sofq}) and (\ref{aofq}). This will here be called the ``local'' method. The second method is to estimate the Bragg intensities and hence the total structure factor using equations (\ref{bhkl}), and (\ref{ihkl}) and (\ref{sbraggint}). This is the ``Bragg'' method. The first method is efficient at short $r$ values but becomes more cumbersome when $r>D_{min}/2$, where $D_{min}$ is the shortest perpendicular distance between any of the three pairs of faces of the parallelpiped that forms the simulation box: beyond that distance it is necessary to take account of the shape of the simulation box when calculating $g_{\alpha\beta}(r)$. The second method captures exactly the long range order implied by the periodic boundary conditions, whatever the shape of the simulation box, but becomes inefficient at high $Q$ due to the large number of Bragg reflections that need to be calculated: this number increases as $Q^{2}$. In practice there is good overlap between the structure factor calculated according to (\ref{sofq}) with that calculated according to (\ref{sbraggint}), which attests to the equivalence of the two formulations. So the question arises how to combine the two approaches so that both the short range local and the long range periodic order are captured accurately \textit{and} efficiently?

A number of ways of achieving this, based mainly on the RMC \cite{rmcpow1999:1,rmcpow2001:1,rmcprofile2007:1} or PDFfit \cite{rmcpow2007:1} modelling hierarchies, have proved successful, and these different methods are described in detail in \cite{rmcprofile2007:1}. An important distinction in EPSR compared to these other methods is that the acceptance or rejection of individual atom and molecule moves is based on the change in the interatomic potential energy function, not the change in fit to the data, and the empirical potential is typically re-calculated only every 5 iterations, where one iteration consists of an attempted move (rotation, translation) of every atom or molecule in the simulation box. Hence in EPSR it is not necessary to calculate either $g_{\alpha\beta}(r)$ or the Bragg peak intensities for each individual atomic or molecular move.

Here we introduce a different method for combining the two approaches to calculating $S(Q)$. The principle is that at short $r$-space distances (large $Q$ structure) the local $g(r)$ method is efficient, while at large distances (small $Q$ structure) the Bragg intensity method is more efficient. To avoid a sharp cut-off for either function, and the associated truncation oscillations that might result, we use a Gaussian merging function of the form
\begin{equation}
m(r) = \exp\left(-\frac{r^2}{2\sigma_r^2}\right), \label{eqmofr}
\end{equation}
where $\sigma_r$ represents the range over which the local structure is regarded as important. Then the total simulated pair distribution function is given by
\begin{equation}
G_{\mathrm{sim}}(r)-1=m(r)\left(G_{\mathrm{local}}(r)-1\right)+(1-m(r))\left(G_{\mathrm{Bragg}}(r)-1\right) \label{eqgrmerge}
\end{equation}
where $G_{\mathrm{local}}(r)$ is obtained directly from the simulation box using (\ref{biggofr}) to some maximum distance $r<D_{min}/2$, and $G_{\mathrm{Bragg}}(r)-1$ is the Fourier transform of $S_{\mathrm{Bragg,int}}(Q)$ (and so exists over all distances in real space):
\begin{equation}
G_{\mathrm{Bragg}}(r)-1=\frac{1}{2\pi^2\rho}\int_0^{\infty}Q^2S_{\mathrm{Bragg,int}}(Q)\frac{\sin Qr}{Qr}dQ, \label{eqgbraggr}
\end{equation}
In the present work $\sigma_r$ is set to 5\AA. (An exception to this rule is the intra-molecular scattering, which is calculated in both $Q$ and $r$ spaces using the ``local'' method for all the atoms and distances in any given molecule.)

As is well known the Fourier transform of a product in $r$ space becomes a convolution in $Q$ space, so if the Gaussian form (\ref{eqmofr}) is used in real space, then the effect is to broaden the intensity at each $Q$ value with a Gaussian function:
\begin{equation}
M(Q,Q^{\prime})=\frac{Q^{\prime}\sigma_r}{Q(2\pi)^\frac{1}{2}}\left[\exp -\frac{(Q^{\prime}-Q)^2\sigma_r^2}{2}-\exp-\frac{(Q^{\prime}+Q)^2\sigma_r^2}{2}\right] \label{eqmofq}
\end{equation}
Hence the simulated structure factor is expressed as
\begin{eqnarray}
S_{\mathrm{sim}}(Q)-1&=&\int M(Q,Q^{\prime})\left( S_{\mathrm{local}}(Q^{\prime})-1 \right) dQ^{\prime} + S_{\mathrm{Bragg,int}}(Q) \nonumber \\
&&- \int M(Q,Q^{\prime}) S_{\mathrm{Bragg,int}}(Q^{\prime}) dQ^{\prime}
\label{eqsofqsim}
\end{eqnarray}
The required convolution (\ref{eqsofqsim}) only needs to be performed when the empirical potential is to be updated, and so does not present a significant additional computational overhead.

\subsection{Running the simulations}

The underlying EPSR methodology is to build a simulation box, run the simulation using some assumed potential parameters, the so-called reference potential (RP), then introduce the empirical potential (EP) to drive the simulated structure factor as close as possible to the measured scattering data. In the present instances, because the simulation box already starts in its proposed final crystalline form, there is little purpose served in running the simulation with the reference potential on its own, since in setting up the polymer (atom-atom spacings, triple atom bond angles, and four-atom dihedral angles) and the unit cell dimensions the measured data have already been used to a large extent to guide the choice of values. For example the position of (100) Bragg reflection is a sensitive indicator of the correct \textbf{a} and \textbf{b} lattice vectors to choose.

The reference potential Lennard-Jones parameters used are listed in Table \ref{tabrp}. They are not based on any previously published values, but were chosen so that the polymer chains would remain reasonably close to their starting positions: if they were allowed to disorder too much, the Bragg peak intensities would decline below their measured values and the empirical potential was often unable to bring the order back because the EP exists only over a finite range in $r$ space, whereas crystal ordering occurs over many unit cells. No Coulomb effective charges are used anywhere in the simulations. In addition some minimum distances between particular atom pairs were used to prevent those particular atom types approaching each other too closely. These are listed in Table \ref{tabmindist}. Once the initial structure refinement was completed the repulsive potential that controls these minimum approach distances was used to hold the pressure of the simulation close to 0.1MPa.

\begin{table}
\caption{\label{tabrp} Lennard-Jones parameters used in the EPSR simulations of PTFE and PCTFE.}
\begin{center}
\item[]\begin{tabular}{@{}llll}
\br
Atom & $\epsilon$ [kJ/mol] & $\sigma$ \AA \\
\mr
C & 0.8 & 3.7 \\
F & 0.3 & 2.0 \\
Cl & 0.3 & 3.3 \\
\br
\end{tabular}
\end{center}
\end{table}
 
\begin{table}
\caption{\label{tabmindist} Minimum distances and (approximate) amplitudes of repulsive potentials used in the EPSR simulations of PTFE and PCTFE.}
\begin{center}
\item[]\begin{tabular}{@{}llll}
\br
Atom pair & Minimum distance [\AA] & Amplitude [kJ/mol] \\
\mr
 (PTFE) \\
C-C & 4.0 & 7 \\
C-F & 2.5 & 0 \\
F-F & 1.8 & 51 \\
\mr
 (PCTFE) \\
C-C & 4.5 & 5 \\
C-F & 2.5 & 2 \\
C-Cl & 3.5 & 31 \\
F-F & 1.8 & 1 \\
F-Cl & 2.5 & 10 \\
Cl-Cl & 3.2 & 44 \\
\br
\end{tabular}
\end{center}
\end{table}

Once the empirical potential was switched on it was allowed to increase in magnitude until there was no discernible improvement in fit to the scattering data. Allowed molecular moves included translations of each polymer molecule independently of its neighbours along all three Cartesian axes, as well as whole molecule rotations about the polymer axis (\textbf{c}-axis). Whole molecule rotations about axes perpendicular to the \textbf{c}-axis were not permitted because these had a highly disordering effect on the structure which could not be recovered. In practice such rotations would be highly unlikely given the length of the real polymers in practice and the corresponding steric hindrance to such rotations. In addition individual atom moves within the polymer occurred on an occasional basis (once every 100 iterations of whole molecule moves) to sample the likely local disorder within the molecules. In addition the empirical potential was included in calculating the energy change of these atom moves to try to obtain the best possible intramolecular fit to the scattering data.

\subsection{Building the polymer strand - PTFE}

\begin{figure*}
\centering
\begin{tabular}{c}
(a)  \\
\includegraphics[width=1\textwidth]{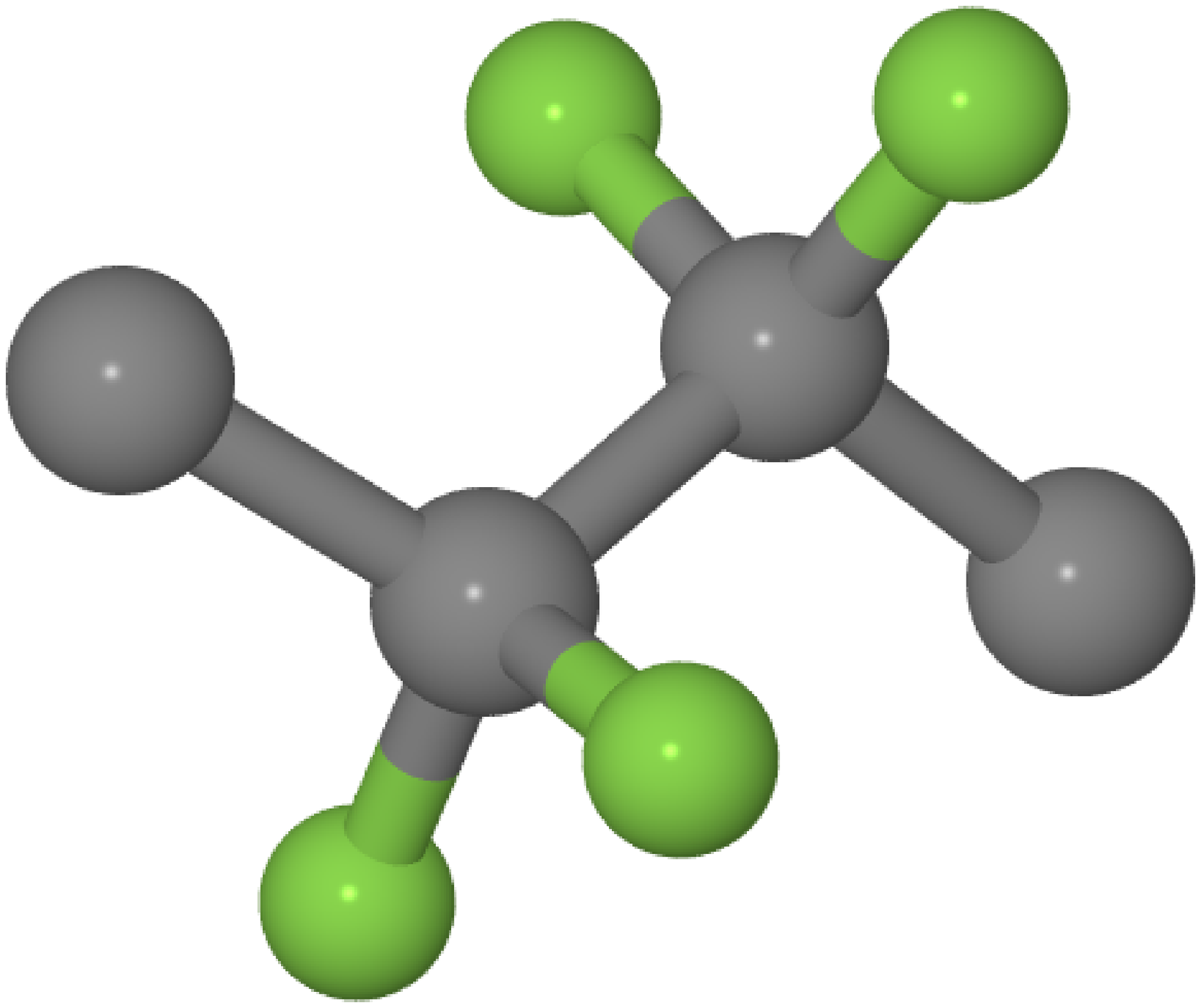} \\
(b) \\
\includegraphics[width=1\textwidth]{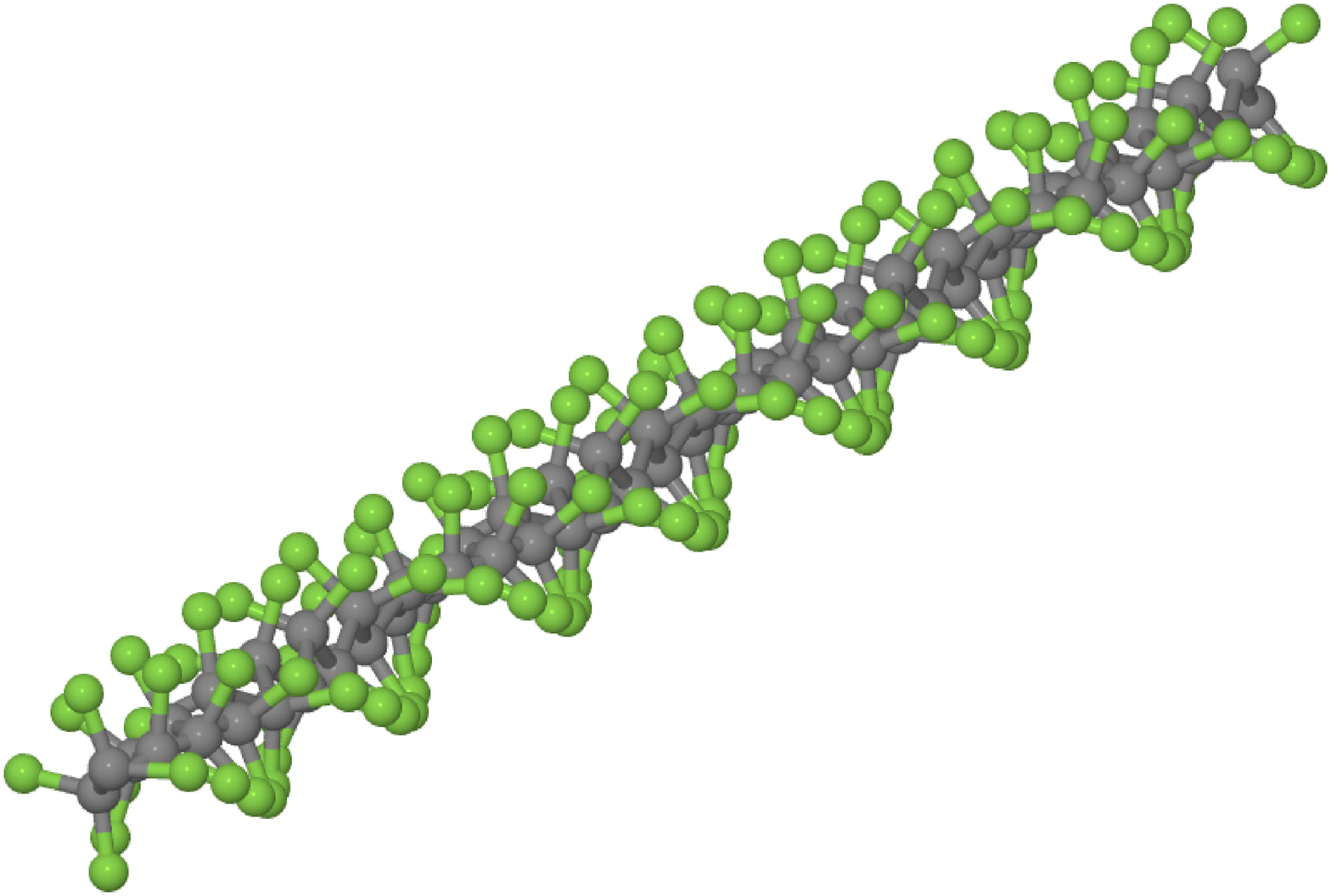} \\
\end{tabular}
\caption{(Colour online) Monomer (a) and oligomer (30 monomers) (b) of PTFE used to construct the EPSR simulations. The monomer shows extra (sacrificial) carbon atoms at each end: these are for the purposes of forming bonds between monomers and are eliminated once the polymer is formed.}
\label{figptfemers} 
\end{figure*}

The construction of the single polymer strand of PTFE was guided by the previous crystallographic results \cite{crystalstruct1954:1,crystalstruct2006:1}, then refined against the new neutron scattering data. The previous work indicated that PTFE has a helical structure which arises from a small twist along each C-C bond away from the exact \textit{trans} conformation ($\phi=180^\circ$), and that the polymers lie parallel to one another, arranged on a hexagonal lattice. The earlier work gives the $a$ and $b$ lattice spacings \cite{crystalstruct1954:1} but is unclear about the $c$ spacing. The later work \cite{crystalstruct2006:1} discusses the near ambient phase transitions that occur in PTFE in much detail, but is remarkable for its lack of numerical values on which to base a model. Neither paper shows any scattering data of the kind presented here, but instead they show single crystal x-ray and electron diffraction images. As a result the present models are something of a guess, based on this earlier work, then refined against the current neutron data. After much trial and error, the best model to emerge was a single strand of helical oligomer which spans the simulation box along the $c$-axis. The internal C-C-C-C dihedral angle, $\phi$, along the polymer backbone is 166$^\circ$, giving a slight twist which leads to a repeat distance of 19.6\AA\ along the $c$-axis and corresponds to 7.5 monomers per repeat distance. This is somewhat larger than the 6.5 value quoted in \cite{crystalstruct1954:1}, but initial attempts to fit the present data with this shorter value, which requires a smaller dihedral angle, were markedly less successful at predicting the Bragg peak positions than the present values. The bond lengths used to define this polymer are C-C: 1.54\AA\ and C-F: 1.36\AA. Triple atom bond angles were defined as F-C-F: 109.47$^{\circ}$ and C-C-C: 116$^{\circ}$, which are the values stated in \cite{crystalstruct1954:1}. Figure \ref{figptfemers} shows examples of the monomer and oligomer consisting of 30 monomers that were constructed in this work. Other oligomers consisted of 45 and 60 monomers were used in separate simulations. The hexagonal lattice constant was set to $a=b=5.687$\AA\, determined from the position of the strong (100) peak, which is to be compared with the value of 5.54\AA\ quoted by \cite{crystalstruct1954:1}. It is not impossible that subtleties of the manufacturing process give rise to these slightly different values from one batch of material to the next. From these axes' dimensions the atomic number density of this model, namely 0.08200 atoms/\AA$^{3}$, corresponds to a macroscopic density of 2.268 gm/cm$^{3}$, which is similar to published values \cite{crystalstruct2006:1}. 

\subsection{Building the polymer strand - PCTFE}

\begin{figure}
\begin{tabular}{c}
(a)  \\
\includegraphics[width=1\textwidth]{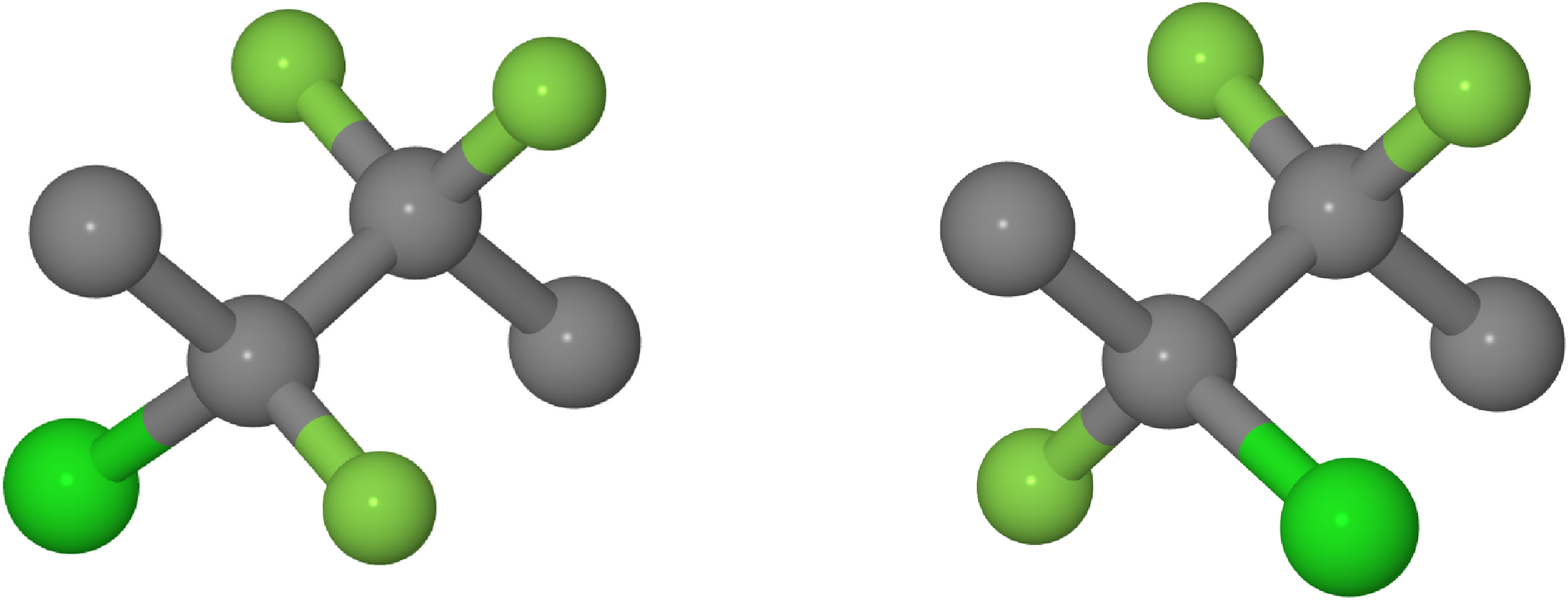} \\
(b) \\
\includegraphics[width=1\textwidth]{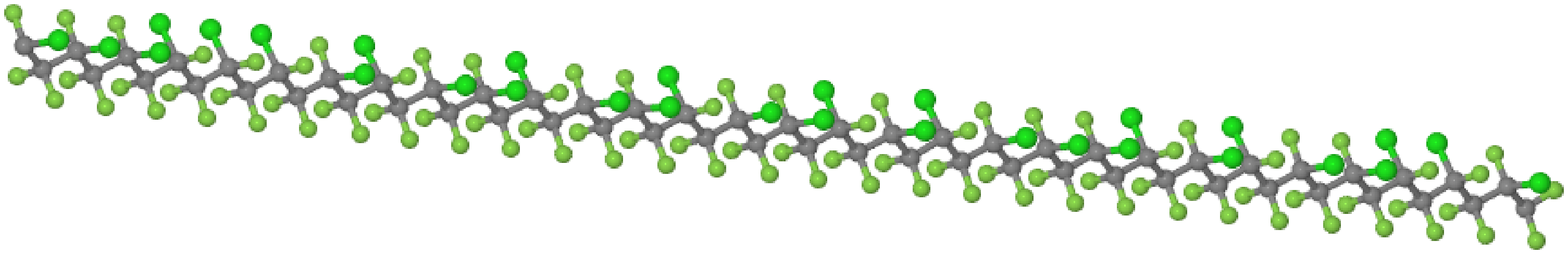}
\end{tabular}
\caption{(Colour online) L (left) and R (right) monomers (a) and oligomer (30 monomers) (b) of PCTFE used to construct the EPSR simulation box. In (b) the monomers have joined together using a random sequence of L and R monomers. As in Fig. \ref{figptfemers} the monomers have extra (sacrificial) carbon atoms at each end: these are for the purposes of forming bonds between monomers and are eliminated once the polymer is formed.}
\label{figpctfemers} 
\end{figure}

The general procedure for building the PCTFE polymer was similar to that described for PTFE. After refinement the C-F bond was the same as PTFE at 1.34\AA\ while the C-C bond was 1.57\AA. The C-Cl bond distance was set to 1.74\AA. The F-C-F bond angle was set to 109.47$^{\circ}$ as for PTFE, while C-C-C angle was set slightly larger than for PTFE at 118$^{\circ}$, and the F-C-Cl bond angle was set to 115.47$^{\circ}$. Initially the C-C-C-C dihedral angle, $\phi$, was set to 166$^{\circ}$ as with PTFE, but it was found a generally better fit could be obtained with $\phi=180^{\circ}$, giving in principle a \textit{trans} conformation zigzag carbon chain, although this became notably disordered after structure refinement. One important difference with PCTFE compared to PTFE is that it is generally thought this polymer is atactic, meaning the Cl atoms can occur on opposite sides of the chain at random \cite{crystalstruct1973:1}. To this end we built two versions of the monomer, labelled ``L'' and ``R'', and formed the oligomer from these units using a random number sequence to determine the order. These units and the resulting chain are shown in Fig. \ref{figpctfemers}. The chosen hexagonal lattice constant at 300K was $a=b=6.374$\AA, based on the position of the (100) Bragg diffraction peak, which compares with the value 6.438\AA\ quoted by \cite{crystalstruct1973:1}. As will be seen shortly there was no information in the data about the possible length of the $c$-axis, so this was set to the length of the 30-mer oligomer, namely $c=79.289$\AA. At 550K the $c$-axis was left unchanged, but the hexagonal lattice constant was set to $a=b=6.481$\AA. The resulting monomers and single oligomer are shown in Fig. \ref{figpctfemers}. In the previous crystallography \cite{crystalstruct1973:1} the polymer was treated as a continuous strand of scattering intensity: the discrete nature of the polymer was not accounted for, whereas in the present work the atomistic nature of the polymer is built into the analysis from the outset. The atomic number density of this model is 0.0653 atoms/\AA$^{3}$ which means a macroscopic density of 2.105 gm/cm$^{3}$. Note that this number density is much lower than for PTFE and no doubt caused by the larger Cl atom creating significantly more disorder in the structure compared to PTFE and requiring more space around it. At 550K the number density drops to 0.06321 atoms/\AA$^{3}$, giving a macroscopic density of 2.036 gm/cm$^{3}$.

\section{Results}

\subsection{PTFE}
\begin{figure}
\centering
\includegraphics[scale=1]{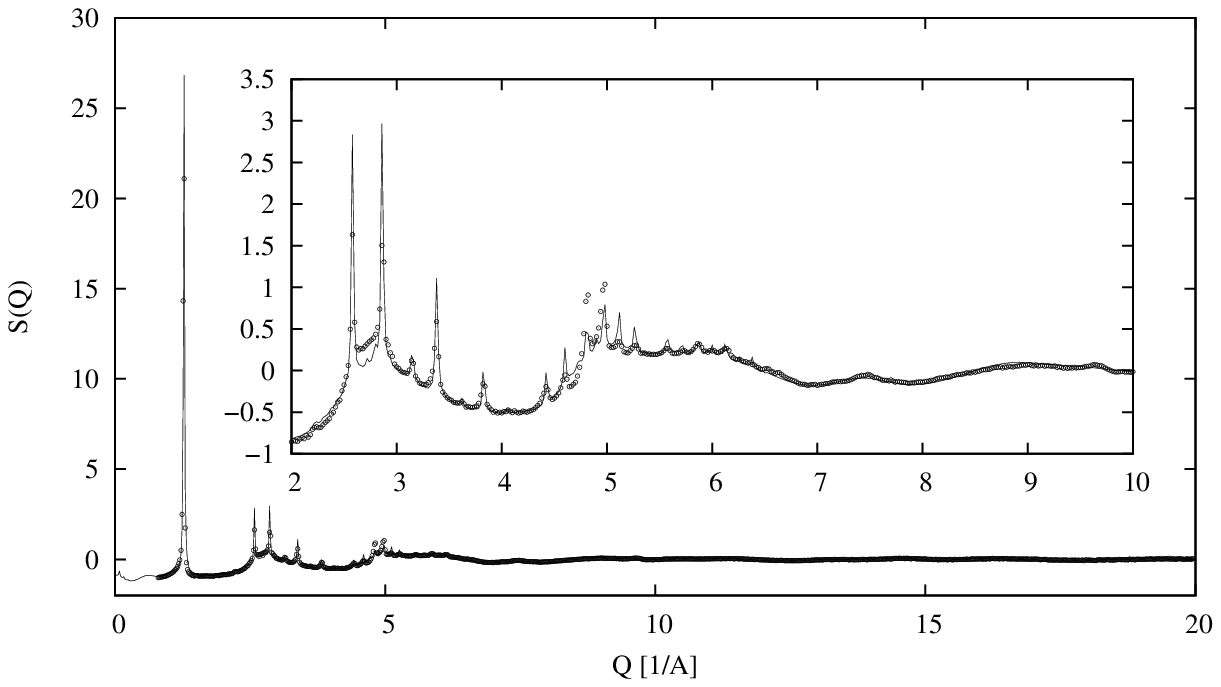}
\caption{EPSR fit (line) to the total scattering data from PTFE (circles) measured on HIPD. The 30-mer oligomer was used for this simulation. The inset shows the region $Q=2-10$\AA$^{-1}$. The Bragg calculation (equation (\ref{sbraggint}) is joined to the local order calculation at $Q=6.5$\AA$^{-1}$ and Lorentzian broadening function (equation (\ref{pofqhkl})) is set to $\gamma=0.01$\AA$^{-1}$.}
\label{figptfe60merfit}
\end{figure}

\begin{figure}
\centering
\includegraphics[scale=1]{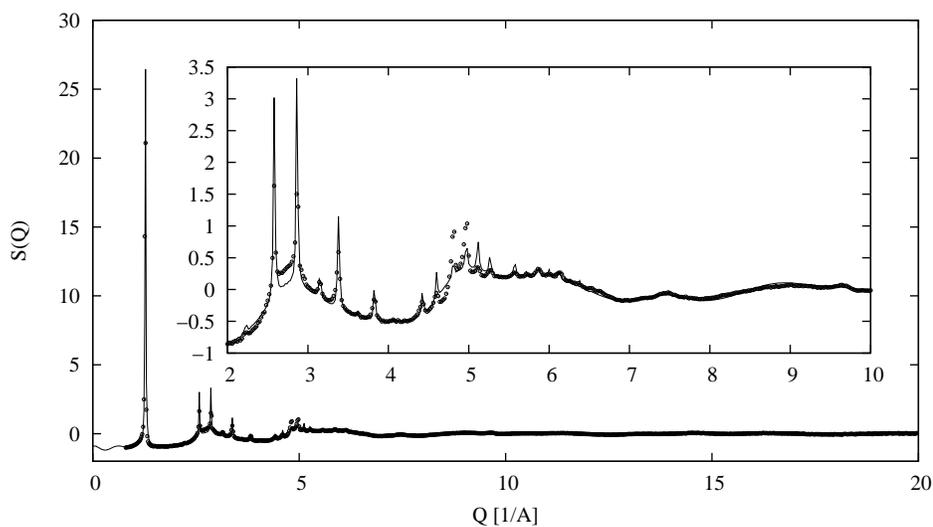}
\caption{This is the same as Fig. \ref{figptfe60merfit} except that the 60-mer oligomer was used for the simulation.}
\label{figptfe120merfit}
\end{figure}

\begin{figure}
\centering
\includegraphics[scale=1]{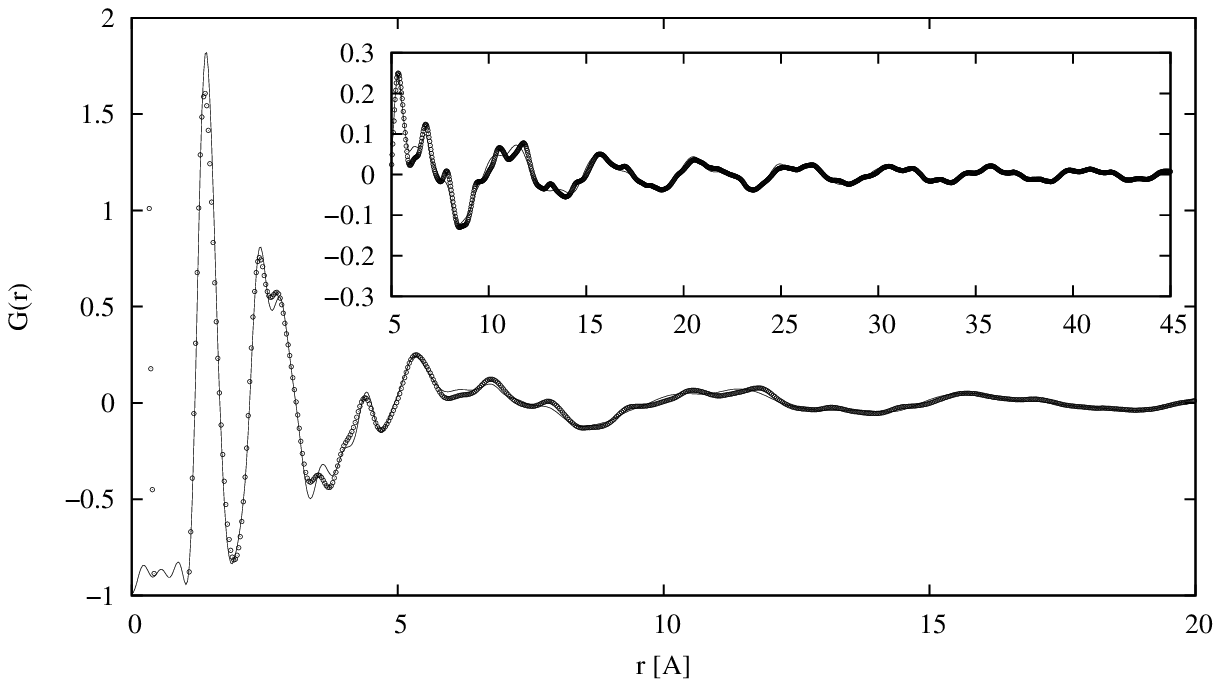}
\caption{Fourier transform of the EPSR fit (line) and total scattering data (circles) for the 60-mer oligomer shown in Fig. \ref{figptfe120merfit}. The inset shows the region $r=5-45$\AA\ in more detail.}
\label{figptfe120merfitrspace}
\end{figure}

For PTFE three EPSR simulations were completed, with 30-, 45-(not shown), and 60-mer oligomers arranged on a hexagonal lattice. Generally the fits were equivalent for all three cases, with a marginally improved fit for the longer oligomers. Figures \ref{figptfe60merfit} and \ref{figptfe120merfit} show the fits with the 30- and 60-mer oligomers respectively in $Q$ space. The equivalent fit in $r$ space for the 60-mer oligomer is shown in Fig. \ref{figptfe120merfit}. Fourier transform of the latter data and fit is shown in Fig. \ref{figptfe120merfitrspace}. As has already been described, the main Bragg diffraction peak near $Q=1.28$\AA$^{-1}$ sets the hexagonal lattice spacing, but the positions of many of the other peaks, in particular the strong peaks near $Q=2.58$\AA$^{-1}$ and $Q=2.86$\AA$^{-1}$ are determined by the repeat distance of the helix: it is the position of these and other peaks that determines the final dihedral angle of $\phi=166^{\circ}$.

Generally the fits in both $Q$ and $r$ space could be regarded as good, bearing in mind that we are attempting to fit a single molecular model to both Bragg intensities and diffuse scattering. The trend in the Bragg peak intensities is mostly well represented by the model, with the obvious exception in the region $Q=4.7-5.0$\AA$^{-1}$ where the observed Bragg peaks are not well represented by the simulation. So far attempts to improve the fit in this region have been unsuccessful. Equally the diffuse scattering is well represented in the model, with some discrepancies near $Q=2.6-2.9$\AA$^{-1}$ and $Q=4.7-5.0$\AA$^{-1}$. In $r$-space these disagreements do not show up as a major problem and it is proposed that the residual discrepancies arise from errors in the assumed internal structure of each oligomer which remain unaccounted for, although how to improve on this is currently unclear. 

\begin{figure}
\centering
\begin{tabular}{c}
(a) \\
\includegraphics[scale=0.5]{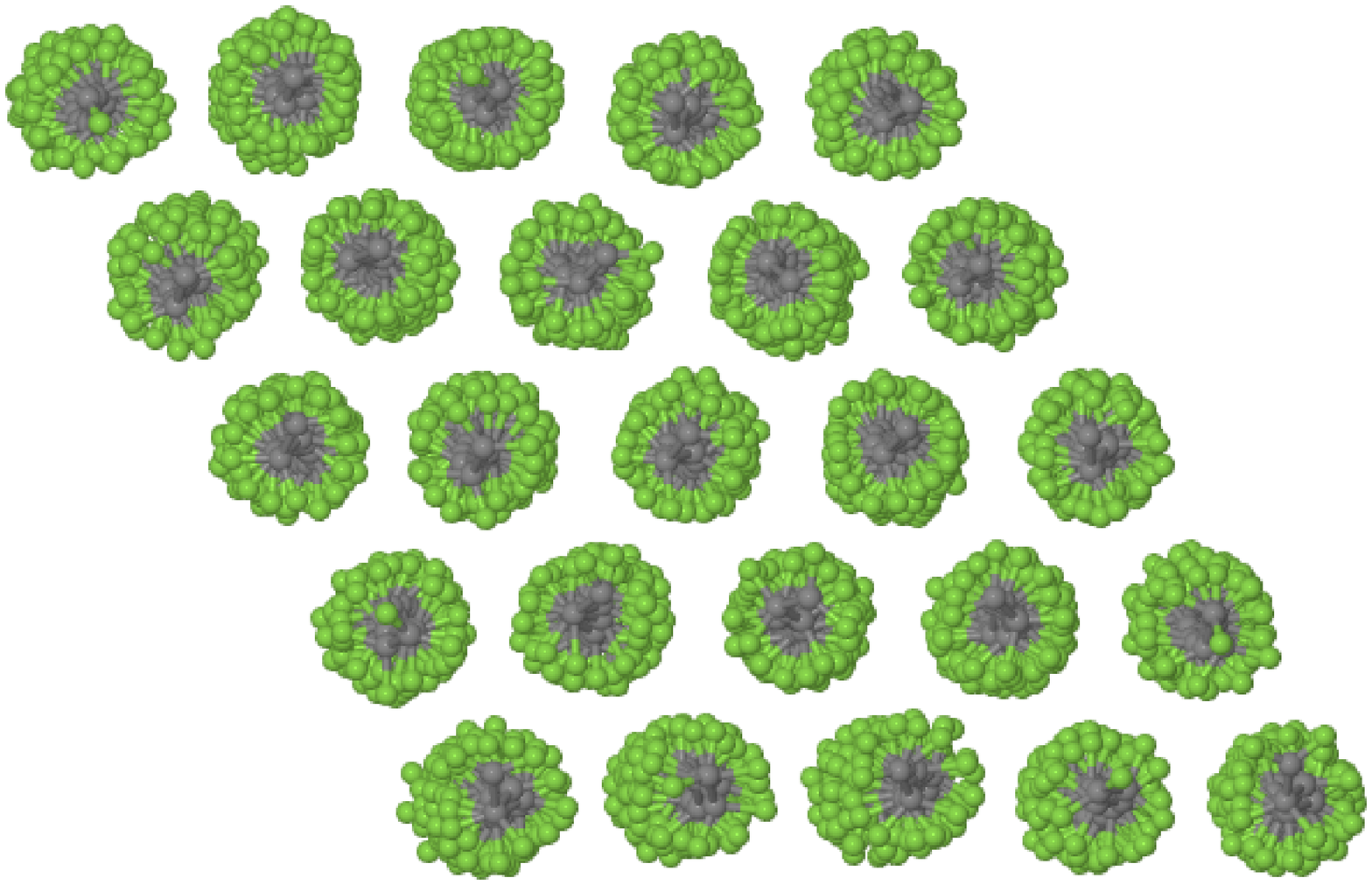} \\
(b) \\
\includegraphics[scale=0.5]{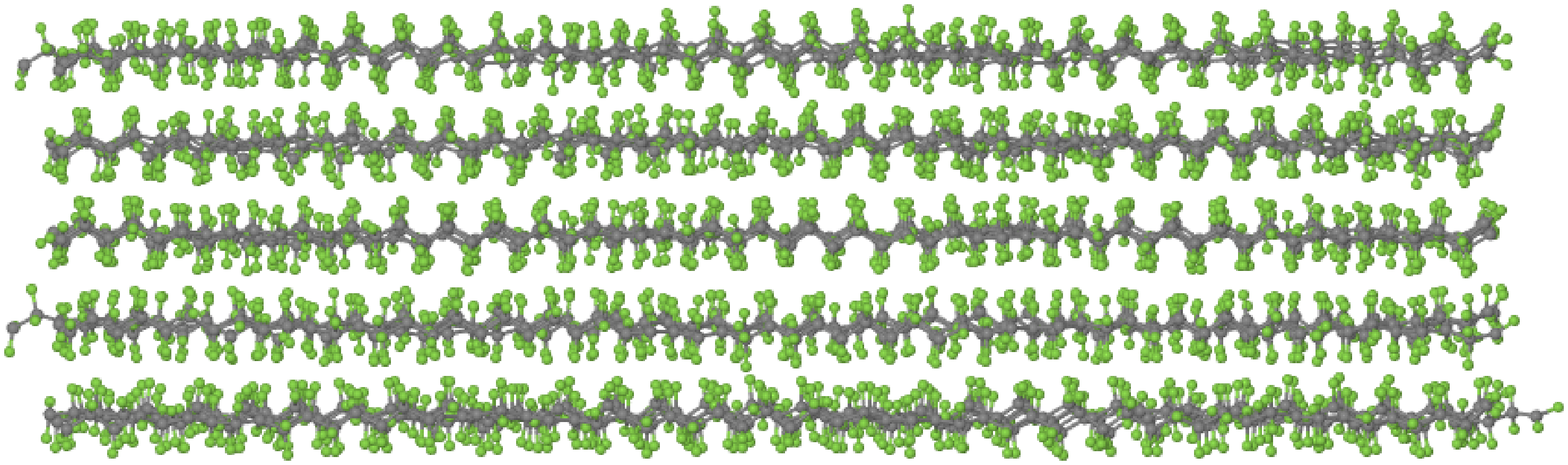} \\
\end{tabular}
\caption{(Colour online) Plots of the EPSR simulation box for PTFE. (a) gives the end-on view, looking down the $c$-axis (= polymer axis), while (b) shows the sideways view along the $a$-axis and perpendicular to the $c$-axis.}
\label{figptfe60merboxplots}
\end{figure}

A view of the simulation box after structure refinement is given in Fig. \ref{figptfe60merboxplots}. It can be seen that looking down the $c$-axis (a) the hexagonal order is disturbed only slightly by the structure refinement although each oligomer has been rotated about this axis and the rotations are not obviously correlated between adjacent polymers. This no doubt is one factor giving rise to the significant diffuse scattering in this material. When viewed at right angles along the $a$-axis (b), however, there is considerable disorder: the repeating nature of the helix is just discernible in these plots, but there does otherwise appear to be significant longitudinal disorder.

\subsection{PCTFE}

\begin{figure}
\centering
\includegraphics[scale=1]{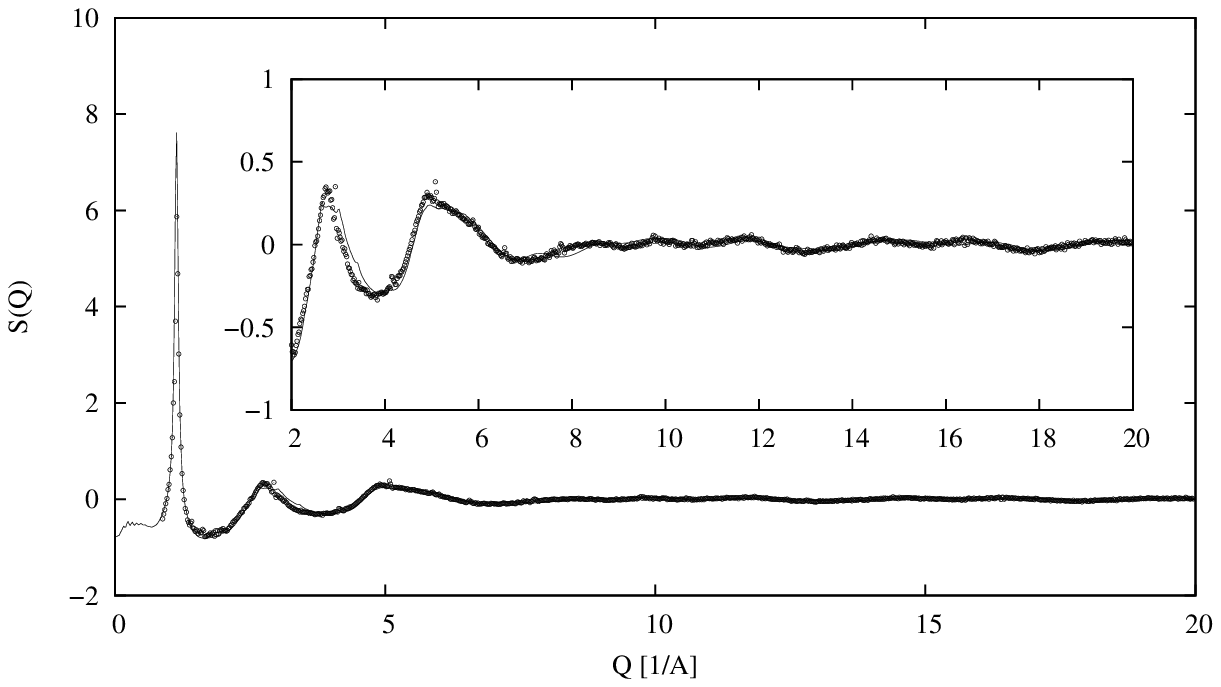}
\caption{EPSR fit (line) to the total scattering data from PCTFE at 300K (circles) measured on NPDF. A 30-mer oligomer was used for this simulation. The inset shows the region $Q=2-20$\AA$^{-1}$ in more detail. The Bragg calculation (equation (\ref{sbraggint}) is joined to the local order calculation at $Q=6.0$\AA$^{-1}$ and the Lorentzian broadening function (equation (\ref{pofqhkl})) is set to $\gamma=0.04$\AA$^{-1}$.}
\label{figpctfe60merfit}
\end{figure}

\begin{figure}
\centering
\includegraphics[scale=1]{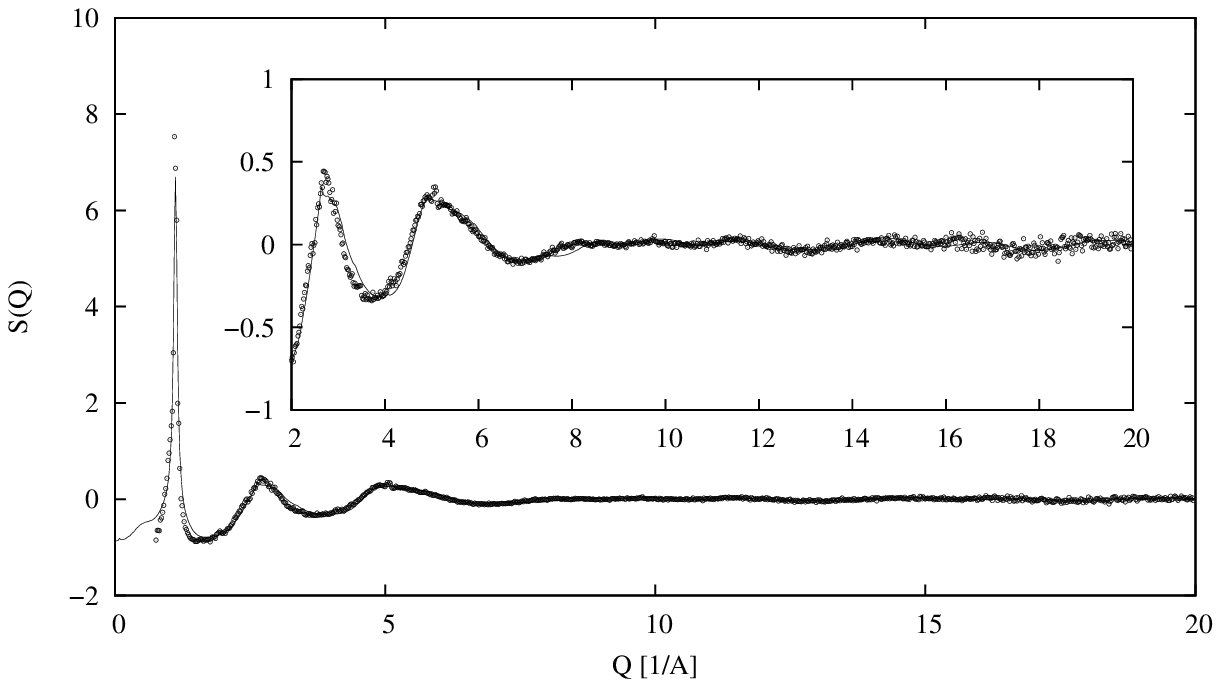}
\caption{EPSR fit (line) to the total scattering data from PCTFE at 550K (circles) measured on HIPD. A 30-mer oligomer was used for this simulation. The inset shows the region $Q=2-20$\AA$^{-1}$ in more detail. The Bragg calculation (equation (\ref{sbraggint}) is joined to the local order calculation at $Q=6.5$\AA$^{-1}$ and the Lorentzian broadening function (equation (\ref{pofqhkl})) is set to $\gamma=0.04$\AA$^{-1}$.}
\label{figpctfe60mer550fit}
\end{figure}

The EPSR fits to the PCTFE total scattering data at 300K (measured on NPDF) and 550K (measured on HIPD) are shown in Figs. \ref{figpctfe60merfit} and \ref{figpctfe60mer550fit} respectively. Comparing these with Figs. \ref{figptfe60merfit} and \ref{figptfe120merfit} two differences are immediately obvious. In the case of PTFE there were many Bragg peaks, while for PCTFE there is only one, the (100) reflection corresponding the hexagonal lattice. At the same time that single Bragg peak in PCTFE is much broader than the equivalent Bragg peak in PTFE. (The value of $\gamma$ in equation (\ref{pofqhkl}) for PTFE was 0.01\AA$^{-1}$ compared to 0.04\AA$^{-1}$ for PCTFE.) Given that the instrumental resolution cannot to be contributing to this difference, since the PCTFE data at 550K and the PTFE data at 300K were both measured on HIPD, this increase in Bragg width can only mean a much smaller particle or domain size for PCTFE compared to PTFE. Since these particle sizes are still much larger than any practical atomistic simulation box, it is necessary to employ the $\gamma$ parameter to capture this effect.

\begin{figure}
\centering
\includegraphics[scale=1]{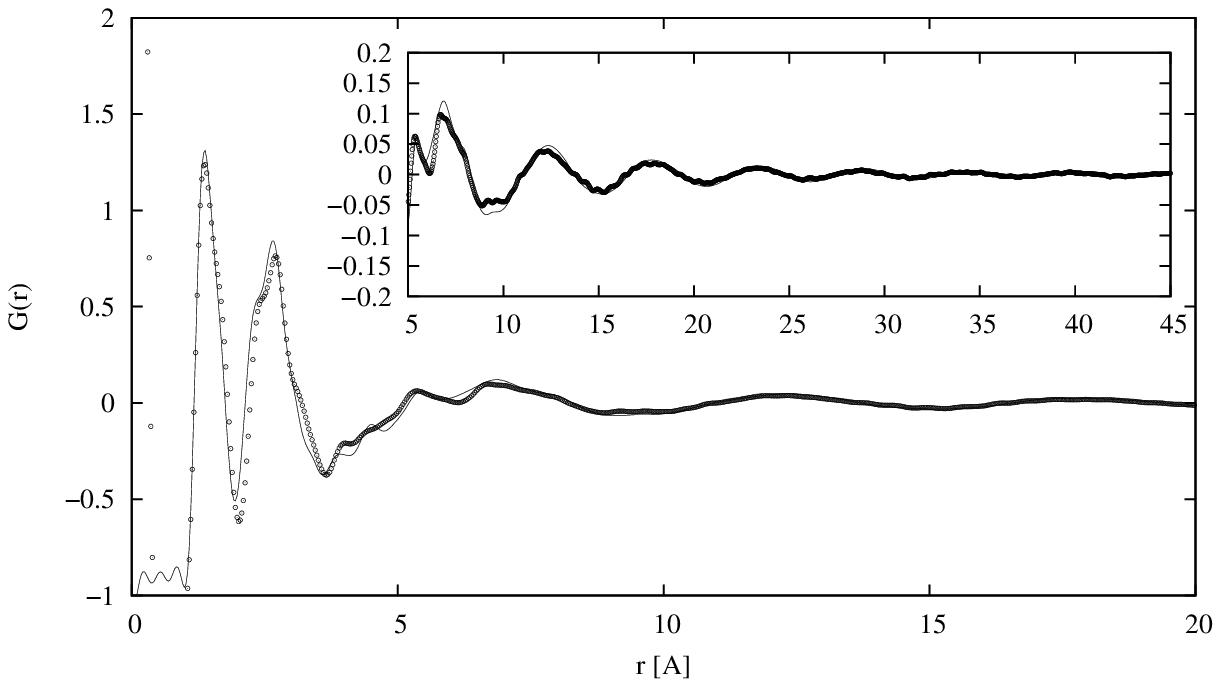}
\caption{Fourier transform of the EPSR fit (line) and total scattering data (circles) for the 30-mer oligomer of PCTFE shown in Fig. \ref{figpctfe60merfit}. The inset shows the region $r=5-45$\AA\ in more detail.}
\label{figpctfe60merfitrspace}
\end{figure}

\begin{figure}
\centering
\begin{tabular}{c}
(a) \\
\includegraphics[scale=0.5]{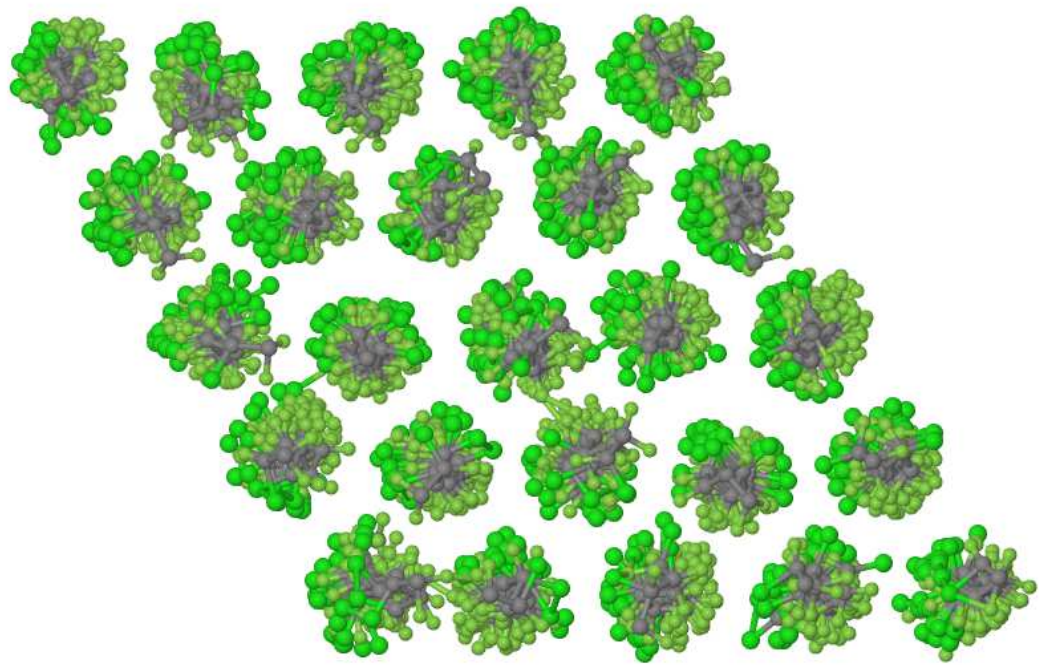} \\
(b) \\
\includegraphics[scale=0.5]{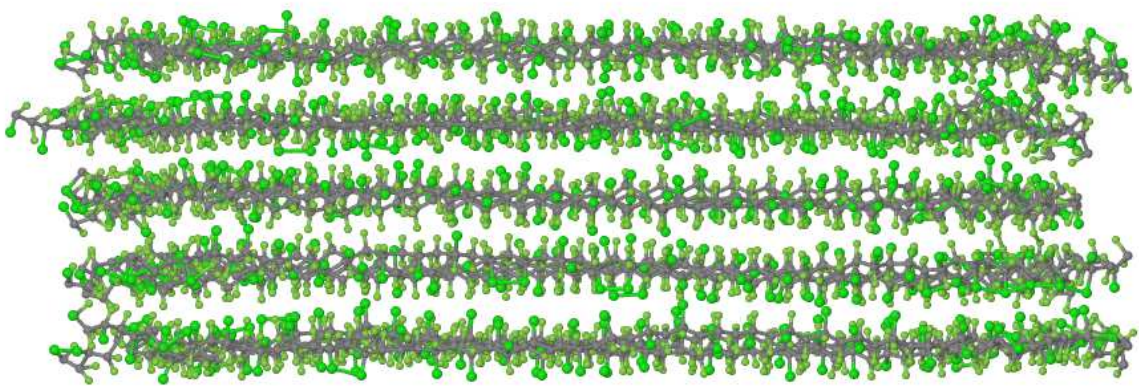} \\
\end{tabular}
\caption{(Colour online) Plots of the EPSR simulation box for PCTFE. (a) gives the end-on view, looking down the $c$-axis (= polymer axis), while (b) shows the sideways view along the $a$-axis and perpendicular to the $c$-axis. Note that in (b) some translational motion of the polymer chains has occurred.}
\label{figpctfe60merboxplots}
\end{figure}

Although the EPSR simulation fits much of the diffuse scattering in the PCTFE data, there are clearly some discrepancies, particularly near $Q=2-5$\AA$^{-1}$. Equally the simulation overestimates the height of the main diffraction peak, and this is seen as a slight mismatch in $r$-space in the amplitude of the large $r$ oscillations, Fig. \ref{figpctfe60merfitrspace}. It seems reasonable to conclude that the current model, while capturing the general features of the data, does not reproduce the structural complexity apparent in the scattering data. The structure refined simulation box is shown in Fig. \ref{figpctfe60merboxplots}, where it is seen that the disorder in both $a-b$ planes and along the $c$-axis is much more marked than with PTFE. At 550K this disorder does not appear to change to any appreciable extent, Fig. \ref{figpctfe60mer550fit}. 

\section{Discussion}

Given its widespread use in chemistry and industry, the crystal structure of polytetrafluoroethylene, PTFE, has been well
studied \cite{ptfestructure1953:1,ptfestructure1981:1,kimmig1994chain,macturk1995role,ptfestructure1999:1,ptfestructure2008:1}, although the original work \cite{crystalstruct1954:1} still carries much weight in the discussion. Also, B. Wunderlich proposed a conformationally disordered crystal (Condis crystal) structure for PTFE from thermal analysis \cite{fai1984thermodynamic}. However, since PTFE does not dissolve or really flow in the melt, the chain microstructure resulting from different processing conditions has not been adequately addressed. From the data shown here clearly PTFE is significantly crystalline at 300K, but we need to be aware that near ambient temperatures it can exist in at least three different phases, II, IV and I \cite{crystalstruct2006:1}: the transition II $\rightarrow$ IV occurs near 292K and the transition IV $\rightarrow$ near 303K, so temperature control is important when accumulating scattering data from this material. The present data have been measured at 300K which means it should be in phase IV, indeed the derived structure is closely analogous to that proposed for this phase. However it is also proposed that chain reversals can occur in this material \cite{kimmig1994chain,crystalstruct2006:1}, where the helix twists the opposite way for part of the chain then reverts back to the original twist. Such chain reversals are not included in the present model: probably a much longer $c$-axis would be needed to capture this effect. Whether these chain reversals would explain why some of the Bragg peak intensities are not reproduced accurately with the present model remains to be seen. 

Another feature of this material to emerge from the literature is that the structure may depend on exactly how it is manufactured, and the manufacture methods have almost certainly changed in the course of time. Such structural differences could explain why a model proposed in 1954 \cite{crystalstruct1954:1} is not quite quantitative against modern scattering data.

In contrast to PTFE, polychlorotrifluoroethylene, PCTFE, is probably encountered less frequently than PTFE, but it is nonetheless quite widespread in industry. The addition of a chlorine atom in place of one fluorine atom gives rise to a markedly lower atomic number density, and introduces much more disorder into the crystalline structure: in the present instance only the main hexagonal peak is visible in the neutron scattering data. There was also significant greater particle size broadening in the particular material used in this study compared to the PTFE sample, and this might be an effect of the chlorine substitution. A previous single crystal study of the structure, \cite{crystalstruct1973:1} suggested the polymer should have a helical structure, whereas here we favour a linear chain structure for the polymer, and the intra-chain correlations are much more disordered compared to PTFE. (Compare Fig. \ref{figpctfe60merboxplots}(b) with Fig. \ref{figptfe60merboxplots}(b).) In fact the best fit was obtained with a non-helical structure. Increasing the temperature to 550K does not appear to cause a significant change in structure other than increase the hexagonal lattice constant slightly.

Generally there is less prior structural information about this material. A relatively recent paper studies the effect of radiation on the structure of PCTFE, \cite{pctfestructure2003:1}, and it does point to an earlier study which claimed to observe degrees of syndiotacticity, isotacticity, and heterotacticity, \cite{pctfestructure1963:1}, suggesting the structure may not in fact be purely atatic as has been assumed here. In fact in the present example, generation of an atactic chain means that some chlorine atoms approach each other at distances of order $\sim 2.2$\AA), that is well inside their normal crystallographic radii. This kind of effect would not be noticed in the earlier x-ray diffraction work \cite{crystalstruct1973:1} which treated the atom scattering as a continuous helix. Based on these observations there may well be more work to be done to understand the structure of this complex material.

It needs to be emphasized that the above quoted intra-chain distances and angles may not be totally reliable. As seen in Figs. \ref{figptfe120merfit}, \ref{figptfe120merfitrspace}, \ref{figpctfe60merfit} and \ref{figpctfe60merfitrspace}, they produce a sensible fit to the scattering data, but this does not imply they are necessarily fully accurate or unique, and it is possible modified values could have produced even better fits. This problem arises here because of the overlap of intra- and inter-chain correlations. When crystallinity is present there is no unique way of disentangling these two contributions in the scattering data, so it is likely resort will have to be made to other measures of bonding, such as NMR spectroscopy, in order to constrain the final values. Hence this situation is distinct from the RMC studies described in the Introduction where it was possible to obtain satisfactory fits from a single chain.

The methods described in the present work are obviously no substitute for the widely used crystallographic methods used to determine polymer structures, see for example \cite{lotz1996structure,andreev2001using}. Indeed for the present methods to work it is probably essential to have a clear idea of the crystal structure before attempting to reconstruct the diffuse scattering. On the other hand modelling both diffuse scattering and crystalline scattering simultaneously provides a useful test of the crystallography without having to resort to ideas of continuous scattering density as have been used in earlier work \cite{crystalstruct1973:1}. There is currently much interest in the crystallisation of polymers from the melt \cite{strobl1996metastable,ryan1997synchrotron,olmsted1998spinodal,schultz2000structural,panine2003combined,strobl2009colloquium}, and it is in these situations where there is likely to be both diffuse and crystalline scattering in the diffraction pattern that methods such as those described here might become useful.

It should also be emphasized that the methods described in this paper should be regarded as complementary to existing procedures for examining local order in crystalline materials, \cite{rmcprofile2007:1,rmcpow2007:1}. In fact the current EPSR approach is designed for studying those crystalline or semi-crystalline systems which include significant - even liquid-like - disorder, such as the study of water in confined geometry, \cite{manc2010:2}. At this stage it remains to be seen whether it will have wider ramifications.

\section{Conclusion}
In the preceding sections a new method of running the empirical potential structure refinement simulation has been introduced. There are three primary advances. One is to build models of long chain polymers. The second is to arrange these in quasi-crystalline lattices and run the simulation on these lattices, as opposed to the disordered materials and liquids that this method has traditionally been applied to \cite{epsrrecs2005:1}. Finally the Bragg peak intensities are calculated exactly based on the simulation box geometry. This proves to be a highly efficient and accurate way to incorporate long range information into the simulated structure factors.

The method was applied to two contrasting polymeric cases. Both involve the polymers being arranged on a hexagonal lattice, but other structural properties are quite different. For PTFE, which is relatively crystalline, although it has a marked degree of diffuse scattering, the Bragg peak positions are well reproduced by the method, although the observed Bragg intensities are not always as accurately reproduced as might be expected. Nonetheless the fits in both $Q$ and $r$ space are very encouraging. For PCTFE, which exhibits considerably more disorder than PTFE, the fits are also highly encouraging, and the remaining small mismatches suggest the assumption of an atactic chain at the outset may require some review. Heating this material from 300K to 550K appears to induce very little change in structure other than a slight increase in the hexagonal lattice parameter. Further work to improve this understanding based on modern scattering data can now be envisaged.

\section*{Acknowledgements}
This work has benefited from the use of the HIPD and NPDF instruments at the Lujan Center at Los Alamos Neutron Science Center, funded by Department of Energy Office of Basic Energy Sciences.  The upgrade of NPDF has been funded by the National Science Foundation through grant DMR00-76488.  Los Alamos National Laboratory is operated by Los Alamos National Security LLC under DOE contract DE-AC52-06NA25396.

\section*{References}

\pagebreak

\pagebreak

\pagebreak

\pagebreak

\end{document}